\def\year{2016}
\documentclass[letterpaper]{article}
\usepackage{aaai}
\usepackage{times}
\usepackage{helvet}
\usepackage{courier}
\usepackage{booktabs}
\usepackage[sort]{natbib}
\usepackage{epsfig}
\usepackage{amssymb}
\usepackage{amsmath}
\usepackage{amsfonts}
\usepackage{multirow}
\usepackage{array}
\usepackage{subcaption}
\usepackage{mathtools}
\usepackage{caption}
\usepackage{url}

\newcommand{\rmd}{VBPR} 
\newcommand{\eq}[1]{Eq.~\ref{#1}}
\newcolumntype{R}[1]{>{\raggedleft\arraybackslash}p{#1}}
\begin{document}
%
\title{\rmd: Visual Bayesian Personalized Ranking from Implicit Feedback}

\author{
\begin{tabular}{*{2}{>{\centering}p{.4\textwidth}}}
Ruining He & Julian McAuley \tabularnewline
UC San Diego & UC San Diego \tabularnewline
\url{r4he@ucsd.edu} & \url{jmcauley@ucsd.edu} 
\end{tabular}
}

\maketitle
\begin{abstract}
\begin{quote}
Modern recommender systems model people and items by discovering or `teasing apart' the underlying dimensions that encode the properties of items and users' preferences toward them. Critically, such dimensions are uncovered based on user feedback, often in \emph{implicit} form (such as purchase histories, browsing logs, etc.); in addition, some recommender systems make use of side information, such as product attributes, temporal information, or review text.
However one important feature that is typically ignored by existing personalized recommendation and ranking methods is the \emph{visual appearance} of the items being considered. In this paper we propose a scalable factorization model to incorporate visual signals into predictors of people's opinions, which we apply to a selection of large, real-world datasets. We make use of visual features extracted from product images using (pre-trained) deep networks, on top of which we learn an additional layer that uncovers the visual dimensions that best explain the variation in people's feedback. This not only leads to significantly more accurate personalized ranking methods, but also helps to alleviate \emph{cold start} issues, and qualitatively to analyze the visual dimensions that influence people's opinions.
\end{quote}
\end{abstract}

\section{Introduction}
Modern Recommender Systems (RSs) provide personalized suggestions by learning from historical feedback and uncovering the preferences of users and the properties of the items they consume. Such systems play a central role in helping people discover items of personal interest from huge corpora, ranging from movies and music \citep{Netflixprize,YahooMusic}, to research articles, news and books \citep{contentCFforNewsRec, GoogleNewsCF}, to tags and even other users \citep{socialRec,tagRec,userRec}.

The `historical feedback' used to train such systems may come in the form of \emph{explicit feedback} such as star ratings, or \emph{implicit feedback} such as purchase histories, bookmarks, browsing logs, search patterns, mouse activities etc. \citep{dwelltime}.
In order to model user feedback in large, real-world datasets, Matrix Factorization (MF) approaches have been proposed to uncover the most relevant latent dimensions in both \emph{explicit} and \emph{implicit} feedback settings~\citep{BellKorSolution, BPR, WRMF, OCCF}. Despite the great success, they suffer from \emph{cold start} issues due to the sparsity of real-world datasets.

\textbf{Visual personalized ranking.} Although a variety of sources of data have been used to build hybrid models to make cold start or context-aware recommendations~\citep{schein}, from text~\citep{topicmf}, to a user's physical location~\citep{socialGeoForEventRec}, to the season or temperature~\citep{Brown1997}, here we are interested in incorporating the \emph{visual appearance} of the items into the preference predictor, a source of data which is typically neglected by existing RSs.
One wouldn't buy a t-shirt from \emph{Amazon} without seeing the item in question, and therefore we argue that this important signal should not be ignored when building a system to recommend such products.

Building on the success of Matrix Factorization methods at uncovering the latent dimensions/factors of people's behavior, our goal here is to ask whether it is possible to uncover the \emph{visual} dimensions that are relevant to people's opinions, and if so,
whether such `visual preference' models shall lead to improved performance at tasks like personalized ranking. Answering these questions requires us to develop scalable methods and representations that are capable of handling millions of user actions, in addition to large volumes of visual data (e.g.~product images) about the content they consume.


In this paper, we develop models that incorporate visual features for the task of personalized ranking on \emph{implicit} feedback datasets. 
By learning the visual dimensions people consider when selecting products we will be able to alleviate \emph{cold start} issues, help explain recommendations in terms of visual signals, and produce personalized rankings that are more consistent with users' preferences. 
Methodologically we model visual aspects of items by using representations of product images derived from a (pre-trained) deep network \citep{Caffe}, on top of which we fit an additional layer that uncovers 
both \emph{visual} and \emph{latent} 
dimensions that are relevant to users' opinions. Although incorporating complex and domain-specific features often requires some amount of manual engineering, we found that visual features are readily available out-of-the-box that are suitable for our task. 

Experimentally our model exhibits significant performance improvements on real-world datasets like \emph{Amazon clothing}, especially when addressing item \emph{cold start} problems. 
Specifically, our main contributions are listed as follows:
\begin{itemize}
\item We introduce a Matrix Factorization approach that incorporates visual signals into predictors of people's opinions while scaling to large datasets. 
\item Derivation and analysis of a Bayesian Personalized Ranking (BPR) based training procedure, which is suitable to uncover visual factors.
\item Experiments on large and novel real-world datasets revealing our method's effectiveness, as well as visualizations of the visual rating space we uncover.
\end{itemize}


\section{Related Work}
Matrix Factorization (MF) methods relate users and items by uncovering latent dimensions such that users have similar representations to items they rate highly, and are the basis of many state-of-the-art 
recommendation approaches.
(e.g.~\citet{BPR, BellKorSolution, Netflixprize}).
When it comes to personalized ranking from \emph{implicit} feedback, traditional MF approaches are challenged by the ambiguity of interpreting `non-observed' feedback.
In recent years, \emph{point-wise} and \emph{pairwise} methods have been successful at adapting MF to address such challenges.

\emph{Point-wise} methods assume non-observed feedback to be inherently negative to some degree. They approximate the task with regression which for each user-item pair predicts its affinity score and then ranks items accordingly. \citet{WRMF} associate different `confidence levels' to positive and non-observed feedback and then factorize the resulting weighted matrix, while \citet{OCCF} sample non-observed feedback as negative instances and factorize a similar weighted matrix.

In contrast to point-wise methods, \emph{pairwise} methods are based on a weaker but possibly more realistic assumption that positive feedback must only be `more preferable' than non-observed feedback.
Such methods directly optimize the ranking of the feedback and are to our knowledge state-of-the-art for implicit feedback datasets. \citet{BPR} propose a generalized Bayesian Personalized Ranking (BPR) framework and experimentally show that BPR-MF (i.e., with MF as the underlying predictor)
outperforms a variety of competitive baselines.
More recently BPR-MF has been extended to accommodate both users' feedback and their social relations \citep{MRBPR, GBPR, ZhaoCIKMSocial}. Our goal here is complementary as we aim to incorporate visual signals into BPR-MF, which presents a quite different set of challenges compared with other sources of data.

Others have developed content-based and hybrid models that make use of a variety of information sources, including text (and context), taxonomies, and user demographics \citep{topicmf,contentCFforNewsRec,taxonomyICDE,socialGeoForEventRec}. However, to our knowledge none of these works have incorporated visual signals into models of users' preferences and uncover visual dimensions as we do here. 

Exploiting visual signals for the purpose of `in-style' image retrieval has been previously proposed. For example, \citet{simo2014neuroaesthetics} predict the fashionability of a person in a photograph and suggest subtle improvements. \citet{StreetFashion} use a street fashion dataset with detailed annotations to identify accessories whose style is consistent with a picture. Another method was proposed by \citet{ClothingSegmentation}, which accepts a query image and uses segmentation to detect clothing classes before retrieving visually similar products from each of the detected classes. \citet{VisualSIGIR} use visual features extracted from CNNs and learn a visual similarity metric to identify visually complementary items to a query image. 
In contrast to our method, the above works focus on visual \emph{retrieval}, which differs from \emph{recommendation} in that such methods aren't personalized to users based on historical feedback, nor do they take into account other factors besides visual dimensions, both of which are essential for a method to be successful at addressing one-class personalized ranking tasks. Thus it is the combination of visual and historical user feedback data that distinguishes our approach from prior work.

\textbf{Visual Features.} Recently, high-level visual features from Deep Convolutional Neural Networks (`Deep CNNs') have seen successes in tasks like object detection \citep{ImageNetchallenge}, photographic style annotations \citep{ImageStyle}, and aesthetic quality categorization \citep{AestheticsDL}, among others. Furthermore, recent transfer learning studies have demonstrated that CNNs trained on one large dataset (e.g.~ImageNet) can be generalized to extract CNN features for other datasets, and outperform state-of-the-art approaches on these new datasets for \emph{different} visual tasks \citep{Decaf, CNNFeature14}. These successes demonstrate the highly generic and descriptive ability of CNN features for visual tasks and persuade us to exploit them for our recommendation task.

\section{\rmd: Visual Bayesian Personalized Ranking}
In this section, we build our visual personalized ranking model (\rmd) to uncover visual and latent (non-visual) dimensions simultaneously.
We first formulate the task in question and introduce our Matrix Factorization based predictor function. Then we develop our training procedure using a Bayesian Personalized Ranking (BPR) framework. The notation we use throughout this paper is summarized in Table \ref{tab:notation}.

\begin{table}
\centering
\caption{Notation \label{tab:notation}}
\begin{tabular}{m{3.29em}l} \toprule
Notation & Explanation\\ \midrule
$\mathcal{U}$, $\mathcal{I}$ & user set, item set\\
$\mathcal{I}_u^+$ & positive item set of user $u$\\
$\widehat{x}_{u,i}$ & predicted `score' user $u$ gives to item $i$\\
$K$ & dimension of latent factors\\
$D$ & dimension of visual factors\\
$F$ & dimension of Deep CNN features\\
$\alpha$ & global offset (scalar)\\
$\beta_u$, $\beta_i$ & user $u$'s bias, item $i$'s bias (scalar)\\
$\gamma_u$, $\gamma_i$ & latent factors of user $u$, item $i$ ($K \times 1$)\\
$\theta_u$, $\theta_i$ & visual factors of user $u$, item $i$ ($D \times 1$)\\
$f_i$ & Deep CNN visual features of item $i$ ($F \times 1$)\\
$\mathbf{E}$ & $D \times F$ embedding matrix\\
$\beta'$ & visual bias vector (visual bias = $\beta'^Tf_i$)\\
\bottomrule
\hline\end{tabular}
\end{table}

\subsection{Problem Formulation} 
Here we focus on scenarios where the ranking has to be learned from users' implicit feedback (e.g.~purchase histories). Letting $\mathcal{U}$ and $\mathcal{I}$ denote the set of users and items respectively, each user $u$ is associated with an item set $\mathcal{I}_u^+$ about which $u$ has expressed explicit positive feedback. In addition, a single image is available for each item $i \in \mathcal{I}$. Using only the above data, our objective is to generate for each user $u$ a personalized ranking of those items about which they haven't yet provided feedback (i.e. $\mathcal{I} \setminus \mathcal{I}_u^+$).

\subsection{Preference Predictor}
Our preference predictor is built on top of Matrix Factorization (MF), which is state-of-the-art for rating prediction as well as modeling implicit feedback, whose basic formulation assumes the following model to predict the preference of a user $u$ toward an item $i$~\citep{korenSurvey}:
\begin{equation} \label{eq:baseline}
\widehat{x}_{u,i} = \alpha + \beta_u + \beta_i + \gamma_u^T \gamma_i,
\end{equation}
where $\alpha$ is global offset, $\beta_u$ and $\beta_i$ are user/item bias terms, and $\gamma_u$ and $\gamma_i$ are $K$-dimensional vectors describing latent factors of user $u$ and item $i$ (respectively). The inner product $\gamma_u^T\gamma_i$ then encodes the `compatibility' between the user $u$ and the item $i$, i.e., the extent to which the user's latent `preferences' are aligned with the products' `properties'.

\begin{figure}
\centering
\includegraphics[width=\columnwidth]{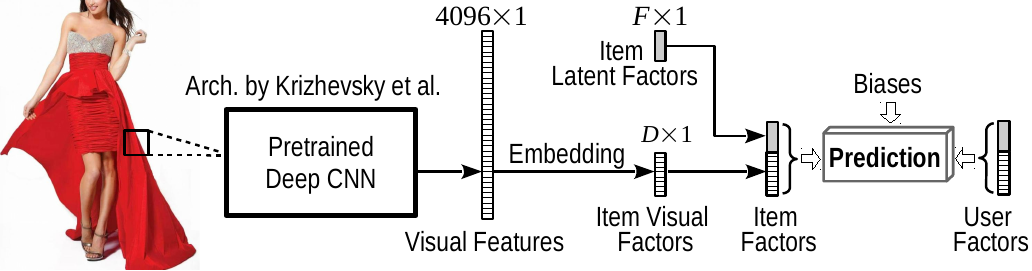}
\caption{Diagram of our preference predictor. Rating dimensions consist of visual factors and latent (non-visual) factors. Inner products between users and item factors model the compatibility between users and items.}
\label{fig:factors}
\end{figure}

Although theoretically latent factors are able to uncover any relevant dimensions, one major problem it suffers from is the existence of `cold' (or `cool') items in the system, about which there are too few associated observations to estimate their latent dimensions. Using explicit features can alleviate this problem by providing an auxiliary signal in such situations. In particular, we propose to partition rating dimensions into visual factors and latent (non-visual) factors, as shown in Figure \ref{fig:factors}. Our extended predictor takes the form
\begin{equation} \label{eq:simple}
\widehat{x}_{u,i} = \alpha + \beta_u + \beta_i + \gamma_u^T \gamma_i + \theta_u^T \theta_i,
\end{equation}
where 
$\alpha$, $\beta$, and $\gamma$ are as in \eq{eq:baseline}. $\theta_u$ and $\theta_i$ are newly introduced $D$-dimensional visual factors whose inner product models the visual interaction between $u$ and $i$, i.e., the extent to which the user $u$ is attracted to each of $D$ visual dimensions. Note that we still use $K$ to represent the number of latent dimensions of our model.

One naive way to implement the above model would be to directly use Deep CNN features $f_i$ of item $i$ as $\theta_i$ in the above equation. However, this would present issues due to the high dimensionality of the features in question, for example the features we use have 4096 dimensions.
Dimensionality reduction techniques like PCA pose a possible solution,
with the potential downside that we would lose much of the expressive power of the original features to explain users' behavior.
Instead, we propose to learn an embedding kernel which linearly transforms such high-dimensional features into a much lower-dimensional (say 20 or so) `visual rating' space:
\begin{equation} \label{eq:embed}
\theta_i = \mathbf{E} f_i
\end{equation}
Here $\mathbf{E}$ is a $D \times F$ matrix embedding Deep CNN feature space ($F$-dimensional) into visual space ($D$-dimensional), where $f_i$ is the original visual feature vector for item $i$. The numerical values of the projected dimensions can then be interpreted as the extent to which an item exhibits a particular visual rating facet. This embedding is efficient in the sense that all items share the same embedding matrix which significantly reduces the number of parameters to learn.

Next, we introduce a visual bias term $\beta'$ whose inner product with $f_i$ models users' overall opinion toward the visual appearance of a given item. In summary, our final prediction model is
\begin{equation} \label{eq:final}
\widehat{x}_{u,i} = \alpha + \beta_u + \beta_i + \gamma_u^T \gamma_i + \theta_u^T (\mathbf{E} f_i) + \beta'^T f_i.
\end{equation}

\subsection{Model Learning Using BPR}
Bayesian Personalized Ranking (BPR) is a pairwise ranking optimization framework which adopts stochastic gradient ascent as the training procedure. A training set $D_S$ consists of triples of the form $(u,i,j)$, where $u$ denotes the user together with an item $i$ about which they expressed positive feedback, and a non-observed item $j$:
\begin{equation}
D_S = \{(u,i,j)|u \in \mathcal{U} \wedge i \in \mathcal{I}_u^+ \wedge j \in \mathcal{I} \setminus \mathcal{I}_u^+ \}.
\end{equation}
Following the notation in \citet{BPR}, $\Theta$ is the parameter vector and $\widehat{x}_{uij}(\Theta)$ denotes an arbitrary function of $\Theta$ that parameterises the relationship between the components of the triple $(u,i,j)$. The following optimization criterion is used for personalized ranking (BPR-OPT):
\begin{equation}
\sum_{(u,i,j) \in D_S} ln \sigma (\widehat{x}_{uij}) - \lambda_{\Theta} ||\Theta||^2
\end{equation}
where $\sigma$ is the logistic (sigmoid) function and $\lambda_{\Theta}$ is a model-specific regularization hyperparameter.

When using Matrix Factorization as the preference predictor (i.e., BPR-MF), $\widehat{x}_{uij}$ is defined as 
\begin{equation} \label{eq:xuij}
\widehat{x}_{uij} = \widehat{x}_{u,i} - \widehat{x}_{u,j},\footnote{Note that $\alpha$ and $\beta_u$ in the preference predictor are canceled out in \eq{eq:xuij}, therefore are removed from the set of parameters.} 
\end{equation}
where $\widehat{x}_{u,i}$ and $\widehat{x}_{u,j}$ are defined by \eq{eq:baseline}. BPR-MF can be learned efficiently using stochastic gradient ascent.
First a triple $(u,i,j)$ is sampled from $D_S$ and then the learning algorithm updates parameters in the following fashion:
\begin{equation}
\Theta \leftarrow \Theta + \eta \cdot (\sigma(-\widehat{x}_{uij}) \frac{\partial \widehat{x}_{uij}}{\partial \Theta} - \lambda_{\Theta}\Theta ), 
\end{equation}
where $\eta$ is the learning rate.

One merit of our model is that it can be learned efficiently using such a sampling procedure with minor adjustments. In our case, $\widehat{x}_{uij}$ is also defined by \eq{eq:xuij} but we instead use \eq{eq:final} as the predictor function for $\widehat{x}_{u,i}$ and $\widehat{x}_{u,j}$ in \eq{eq:xuij}. Compared to BPR-MF, there are now two sets of parameters to be updated: (a) the non-visual parameters, and (b) the newly-introduced visual parameters. Non-visual parameters can be updated in the same form as BPR-MF (therefore are suppressed for brevity), while visual parameters are updated according to:
\begin{eqnarray*}
\theta_u \leftarrow \theta_u + \eta \cdot (\sigma(-\widehat{x}_{uij})  \mathbf{E}(f_i - f_j) - \lambda_{\Theta}  \theta_u ),\\
\beta' \leftarrow \beta' + \eta \cdot (\sigma(-\widehat{x}_{uij}) (f_i - f_j) -  \lambda_{\beta} \beta' ),\\
\mathbf{E} \leftarrow \mathbf{E} + \eta \cdot (\sigma(-\widehat{x}_{uij}) \theta_u (f_i - f_j)^T - \lambda_{\mathbf{E}}  \mathbf{E} ).
\end{eqnarray*}

Note that our method introduces an additional hyperparameter $\lambda_{\mathbf{E}}$ to regularize the embedding matrix $\mathbf{E}$. 
We sample users uniformly to optimize the average AUC across all users to be described in detail later. All hyperparameters are tuned using a validation set as we describe in our experimental section later.

\subsection{Scalability}
The efficiency of the underlying BPR-MF makes our models similarly scalable. Specifically, BPR-MF requires $\mathcal{O}(K)$ to finish updating the parameters for each sampled triple $(u,i,j)$. In our case we need to update the visual parameters as well. In particular, updating $\theta_u$ takes $\mathcal{O}(D \times F) = \mathcal{O}(D)$, $\beta'$ takes $\mathcal{O}(F)$, and $\mathbf{E}$ takes $\mathcal{O}(D \times F) = \mathcal{O}(D)$, where $F$ is the dimension of CNN features (fixed to 4096 in our case). Therefore the total time complexity of our model for updating each triple is $\mathcal{O}(K + D)$ (i.e. $\mathcal{O}(K) + \mathcal{O}(D \times F)$), i.e., linear in the number of dimensions. Note that visual feature vectors ($f_i$) from Deep CNNs are sparse, which significantly reduces the above worst-case running time.

\section{Experiments}
In this section, we perform experiments on multiple real-world datasets. These datasets include a variety of settings where visual appearance is expected to play a role in consumers' decision-making process.

\begin{table}
\centering
\renewcommand{\tabcolsep}{6pt}
\caption{Dataset statistics (after preprocessing)} 
\begin{tabular}{lrrr} \toprule
Dataset                  &\#users    & \#items     &\#feedback   \\ \midrule
\emph{Amazon Women}      & 99,748	 & 331,173     & 854,211	 \\
\emph{Amazon Men}        & 34,212    & 100,654     & 260,352     \\
\emph{Amazon Phones}     & 113,900   & 192,085     & 964,477     \\
\emph{Tradsy.com} 		 & 19,823    & 166,526     & 410,186     \\ \midrule
Total                    & 267,683   & 790,438     & 2,489,226   \\ \bottomrule
\hline\end{tabular}
\label{table:dataset}
\end{table}

\begin{table*}
\centering
\caption{AUC on the test set $\mathcal{T}$ (\#factors = 20). The best performing method on each dataset is boldfaced.}
\begin{tabular}{llccccccrr} \toprule
\multirow{2}{*}{Dataset}  &\multirow{2}{*}{Setting}   &(a)    &(b)    &(c)    &(d)    &(e)    &(f)             & \multicolumn{2}{c}{improvement}\\ 
                          &                           &RAND   &MP     &IBR    &MM-MF  &BPR-MF &\rmd            &f vs. best &f vs. e \\ \midrule
\multirow{2}{*}{\emph{Amazon Women}}     &All Items   &0.4997 &0.5772 &0.7163 &0.7127 &0.7020 &\textbf{0.7834} &9.4\%      &11.6\%   \\
                                  &\emph{Cold Start}  &0.5031 &0.3159 &0.6673 &0.5489 &0.5281 &\textbf{0.6813} &2.1\%      &29.0\%   \\[4pt]
\multirow{2}{*}{\emph{Amazon Men}}       &All Items   &0.4992 &0.5726 &0.7185 &0.7179 &0.7100 &\textbf{0.7841} &9.1\%      &10.4\%   \\
                                  &\emph{Cold Start}  &0.4986 &0.3214 &0.6787 &0.5666 &0.5512 &\textbf{0.6898} &1.6\%      &25.1\%   \\[4pt]
\multirow{2}{*}{\emph{Amazon Phones}}    &All Items   &0.5063 &0.7163 &0.7397 &0.7956 &0.7918 &\textbf{0.8052} &1.2\%      &1.7\%    \\
                                  &\emph{Cold Start}  &0.5014 &0.3393 &\textbf{0.6319} &0.5570 &0.5346 &0.6056 &-4.2\%     &13.3\%   \\[4pt]             
\multirow{2}{*}{\emph{Tradesy.com}}      &All Items   &0.5003 &0.5085 &N/A    &0.6097 &0.6198 &\textbf{0.7829} &26.3\%     &26.3\%   \\
                                  &\emph{Cold Start}  &0.4972 &0.3721 &N/A    &0.5172 &0.5241 &\textbf{0.7594} &44.9\%     &44.9\%   \\ \bottomrule
                                      
\hline\end{tabular}
\label{table:aucres}
\end{table*}

\subsection{Datasets}
The first group of datasets are from \emph{Amazon.com} introduced by \citet{VisualSIGIR}. We consider two large categories where visual features have already been demonstrated to be meaningful, namely Women's and Men's Clothing.
We also consider Cell Phones \& Accessories, where we expect visual characteristics to play a smaller but possibly still significant role.
We take users' review histories as implicit feedback and use one image per item to extract visual features.

We also introduce a new dataset from \emph{Tradesy.com}, a second-hand clothing trading community. It discloses users' purchase histories and `thumbs-up', which we use together as positive feedback. Note that recommendation in this setting inherently involves \emph{cold start} prediction due to the `one-off' trading characteristic of second-hand markets. Thus to design a meaningful recommender system for such a dataset it is critical that visual information be considered.

We process each dataset by extracting implicit feedback and visual features as already described. We discard users $u$ where $|\mathcal{I}_u^+| < 5$.
Table~\ref{table:dataset} shows statistics of our datasets,
all of which shall be made available at publication time.

\subsection{Visual Features}
For each item $i$ in the above datasets, we collect one product image and extract visual features $f_i$ using the Caffe reference model \citep{Caffe}, which implements the CNN architecture proposed by~\citet{DeepCNNArchitecture}. The architecture has 5 convolutional layers followed by 3 fully-connected layers, and has been pre-trained on 1.2 million ImageNet (ILSVRC2010) images. In our experiments, we take the output of the second fully-connected layer (i.e. FC7), to obtain an $F = 4096$ dimensional visual feature vector $f_i$. 

\subsection{Evaluation Methodology}
We split our data into training/validation/test sets by selecting for each user $u$ a random item to be used for validation $\mathcal{V}_u$ and another for testing $\mathcal{T}_u$. All remaining data is used for training.
The predicted ranking is evaluated on $\mathcal{T}_u$ with the widely used metric AUC (\emph{Area Under the ROC curve}):
\begin{equation}
\mathit{AUC} =  \frac{1}{|\mathcal{U}|}  \sum_u   \frac{1}{|E(u)|}   \sum_{(i,j) \in E(u)}  \delta (\widehat{x}_{u,i} > \widehat{x}_{u,j})
\end{equation}
where the set of evaluation pairs for user $u$ is defined as
\begin{equation}
E(u) = \{(i,j) | (u,i) \in \mathcal{T}_u \wedge (u,j) \notin  (\mathcal{P}_u \cup \mathcal{V}_u \cup \mathcal{T}_u) \},
\end{equation}
and $\delta(b)$ is an indicator function that returns 1 iff $b$ is $\mathit{true}$.
In all cases we report the performance on the test set $\mathcal{T}$ for the hyperparameters that led to the best performance on the validation set $\mathcal{V}$.

\subsection{Baselines}
Matrix Factorization (MF) methods are known to have state-of-the-art performance for implicit feedback datasets. Since there are no comparable visual-aware MF methods, we mainly compare against state-of-the-art MF models, in addition to a recently proposed content-based method.
\begin{itemize} 
\item \textbf{Random (RAND):} This baseline ranks items randomly for all users.
\item \textbf{Most Popular (MP):} This baseline ranks items according to their popularity and is non-personalized.
\item \textbf{MM-MF:} A pairwise MF model from \citet{MyMediaLite}, which is optimized for a hinge ranking loss on $x_{uij}$ and trained using SGA as in BPR-MF.
\item \textbf{BPR-MF:} This pairwise method was introduced by \citet{BPR} and is the state-of-the-art of personalized ranking for \emph{implicit} feedback datasets.
\end{itemize}

We also include a `content-based' baseline for comparison against another method which makes use of visual data, though which differs in terms of problem setting and data (it does not make use of feedback but rather graphs encoding relationships between items as input):
\begin{itemize} 
\item \textbf{Image-based Recommendation (IBR):} Introduced by \citet{VisualSIGIR}, it learns a visual space and retrieves stylistically similar items to a query image. Prediction is then performed by nearest-neighbor search in the learned visual space.
\end{itemize}

Though significantly different from the pairwise methods considered above, for comparison we also compared against a \emph{point-wise} method, \textbf{WRMF} \citep{WRMF}.

Most baselines are from MyMediaLite \citep{MyMediaLite}. For fair comparison, we use the same total number of dimensions for all MF based methods. In our model, visual and non-visual dimensions are fixed to a fifty-fifty split for simplicity, though further tuning may yield better performance. All experiments were performed on a standard desktop machine with 4 physical cores and 32GB main memory.

\noindent \textbf{Reproducibility.}
All hyperparameters are tuned to perform the best on the validation set. On \emph{Amazon}, regularization hyperparamter $\lambda_{\Theta} = 10$ works the best for BPR-MF, MM-MF and \rmd~in most cases. While on \emph{Tradesy.com}, $\lambda_{\Theta}=0.1$ is set for BPR-MF and \rmd~and $\lambda_{\Theta}=1$ for MM-MF.
$\lambda_{\mathbf{E}}$ is always set to $0$ for \rmd. For IBR, the rank of the Mahalanobis transform is set to 100, which is reported to perform very well on \emph{Amazon} data. All of our code and datasets shall be made available at publication time so that our experimental evaluation is completely reproducible.

\subsection{Performance}
Results in terms of the average AUC on different datasets are shown in Table \ref{table:aucres} (all with 20 total factors). For each dataset, we report the average AUC on the full test set $\mathcal{T}$ (denoted by `All Items'), as well as a subset of $\mathcal{T}$ which only consists of items that had fewer than five positive feedback instances in the training set (i.e., \emph{cold start}). These cold start items account for around 60\% of the test set for the two \emph{Amazon} datasets, and 80\% for \emph{Tradesy.com}; this means for such sparse real-world datasets, a model must address the inherent \emph{cold start} nature of the problem and recommend items accurately in order to achieve acceptable performance. The main findings from Table \ref{table:aucres} are summarized as follows:

\begin{figure}
\centering
\includegraphics[width=\columnwidth]{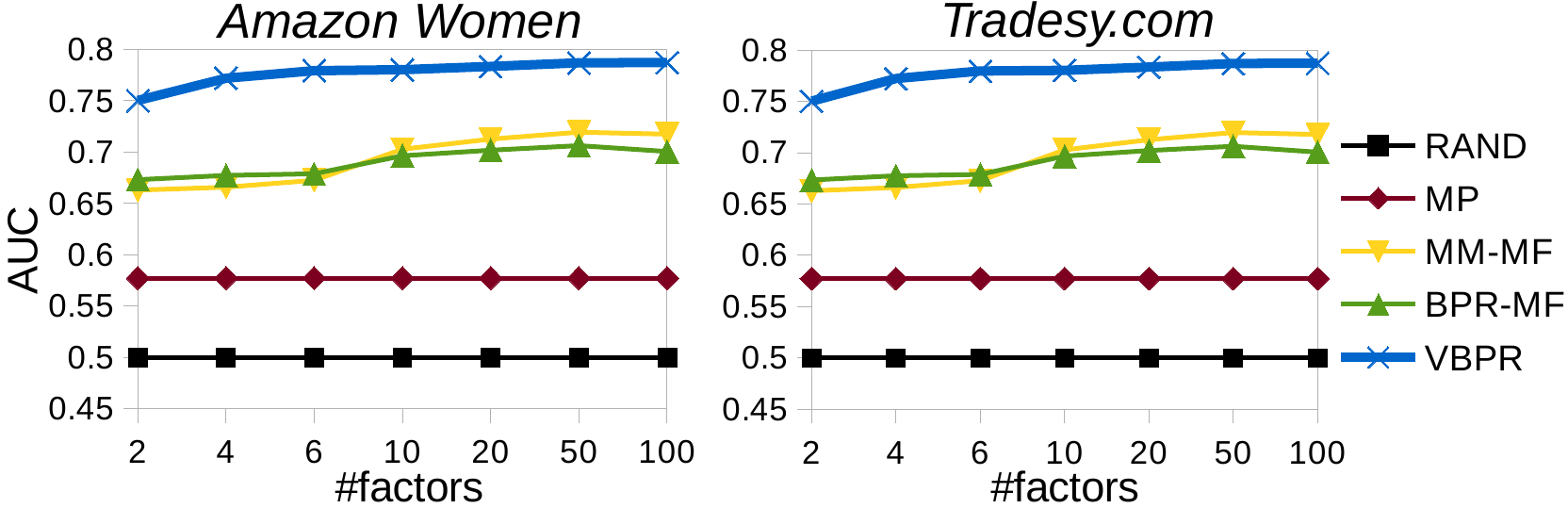}
\caption{AUC with varying dimensions.}
\label{fig:AUCFactor}
\end{figure}
\begin{figure}
\centering
\includegraphics[width=\columnwidth]{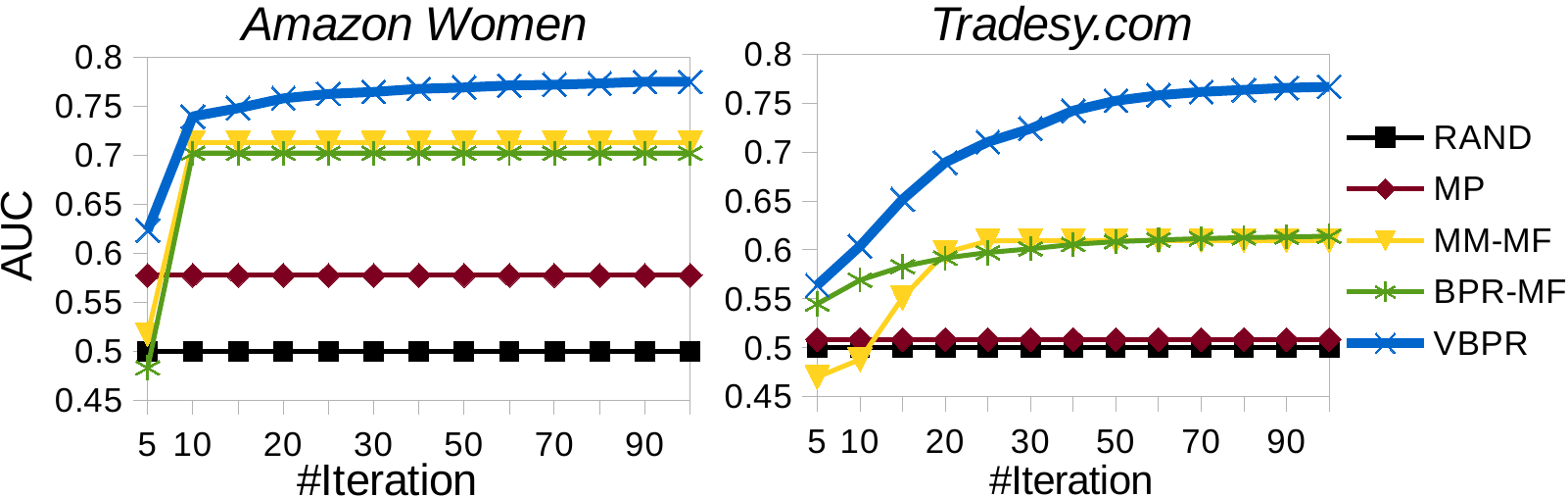}
\caption{AUC with training iterations (\#factors=20).}
\label{fig:AUCIter}
\end{figure}
\begin{figure*}
\centering
\includegraphics[width=\textwidth]{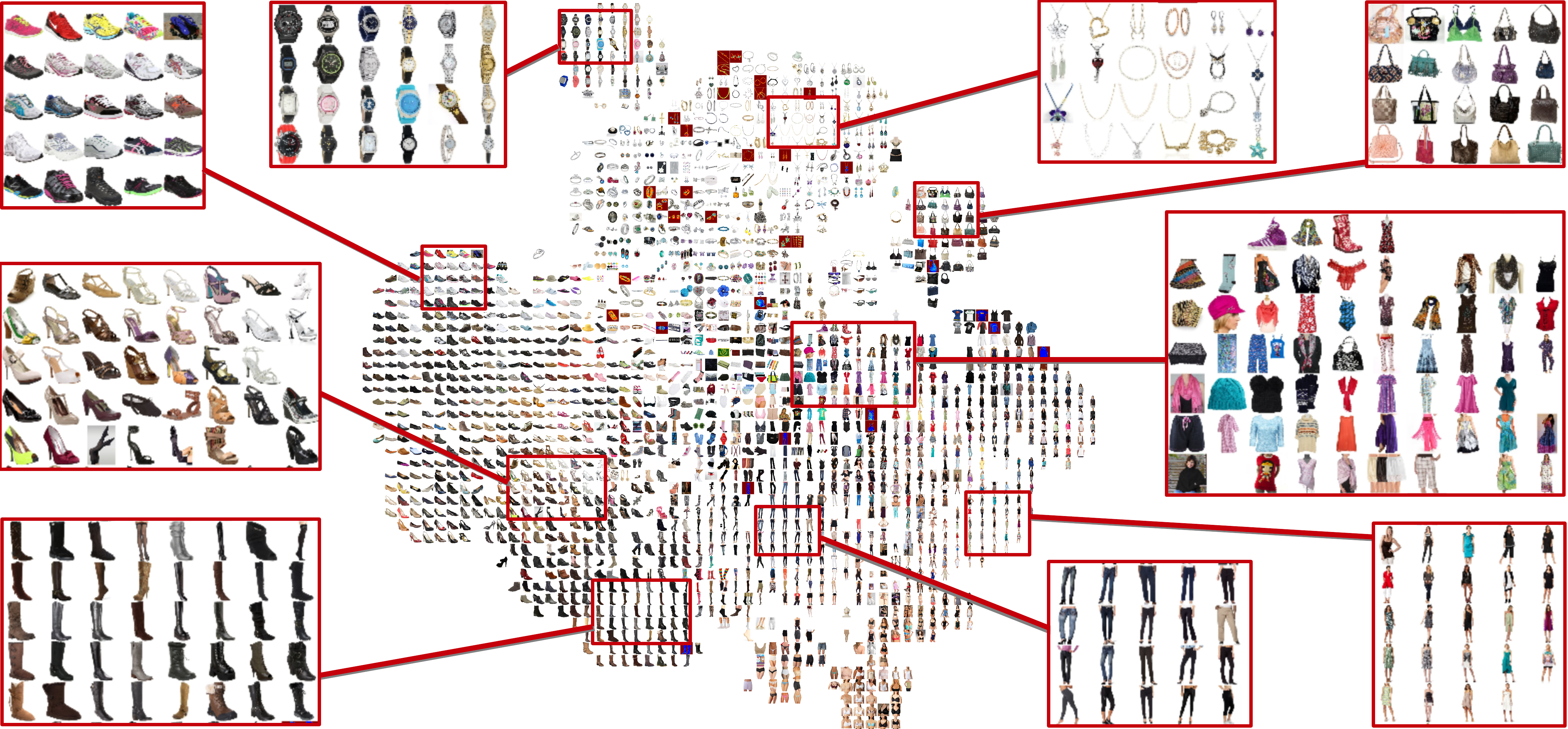}
\caption{2-D visualization (with t-SNE \citep{tsne}) of the 10-D visual space learned from \emph{Amazon Women}. All images are from the test set. For clarity, the space is discretized into a grid and for each grid cell one image is randomly selected among overlapping instances.}
\label{fig:visualspace}
\end{figure*}


\begin{enumerate}
  \item Building on top of BPR-MF, \rmd~on average improves on BPR-MF by over 12\% for all items, and more than 28\% for \emph{cold start}. This demonstrates the significant benefits of incorporating CNN features into our ranking task.
  \item As expected, IBR outperforms BPR-MF \& MM-MF in \emph{cold start} settings where pure MF methods have trouble learning meaningful factors. Moreover, IBR loses to MF methods for \emph{warm start} since it is not trained on historical user feedback.
  \item By combining the strengths of both MF and content-based methods, \rmd~outperforms all baselines in most cases.
  \item Our method exhibits particularly large improvements on \emph{Tradesy.com}, since it is an inherently \emph{cold start} dataset due to the `one-off' nature of trades.
  \item Visual features show greater benefits on clothing than cellphone datasets. Presumably this is because visual factors play a smaller (though still significant) role when selecting cellphones as compared to clothing.
  \item Popularity-based methods are particularly ineffective here, as cold items are inherently `unpopular'.
\end{enumerate}

Finally, we found that \emph{pairwise} methods indeed outperform \emph{point-wise} methods (WRMF in our case) on our datasets, consistent with our analysis in Related Work. We found that on average, \rmd~beats WRMF by 14.3\% for all items and 20.3\% for \emph{cold start} items.


\noindent\textbf{Sensitivity.}
As shown in Figure \ref{fig:AUCFactor}, MM-MF, BPR-MF, and \rmd~perform better as the number of factors increases, which demonstrates the ability of pairwise methods to avoid overfitting.
Results for other \emph{Amazon} categories are similar and suppressed for brevity.

\noindent\textbf{Training Efficiency.}
In Figure \ref{fig:AUCIter} we demonstrate the AUC (on the test set) with increasing training iterations. Generally speaking, our proposed model takes longer to converge than MM-MF and BPR-MF, though still requires only around 3.5 hours to train to convergence on our largest dataset (Women's Clothing).

\subsection{Visualizing Visual Space}
\rmd~maps items to a low-dimensional `visual space,' such that items with similar styles (in terms of how users evaluate them) are mapped to nearby locations. We visualize this space (for Women's Clothing) in Figure \ref{fig:visualspace}. We make the following two observations: (1) although our visual features are extracted from a CNN pre-trained on a different dataset, by using the embedding we are nevertheless able to learn a `visual' transition (loosely) across different subcategories, which confirms the expressive power
of the extracted features; and (2) \rmd~not only helps learn the hidden taxonomy, but also more importantly discovers the most relevant underlying visual dimensions and maps items and users into the uncovered space.

\section{Conclusion \& Future Work}
Visual decision factors influence many of the choices people make, from the clothes they wear to their interactions with each other. In this paper, we investigated the usefulness of visual features for personalized ranking tasks on \emph{implicit feedback} datasets. We proposed a scalable method that incorporates visual features extracted from product images into Matrix Factorization, in order to uncover the `visual dimensions' that most influence people's behavior. Our model is trained with Bayesian Personalized Ranking (BPR) using stochastic gradient ascent. Experimental results on multiple large real-world datasets demonstrate that we can significantly outperform state-of-the-art ranking techniques and alleviate \emph{cold start} issues.

As part of future work, we will further extend our model with temporal dynamics to account for the drifting of fashion tastes over time. Additionally, we are also interested in investigating the efficacy of our proposed method in the setting of \emph{explicit} feedback.


\setlength{\bibhang}{0pt}
\bibliographystyle{aaai}
\bibliography{visual}

\end{document}